\documentstyle[aps,pre,floats,amsmath,amssymb,graphicx]{revtex}
\def\smallcaption#1{\caption{\footnotesize #1}}
\begin{document}
 
\renewcommand{\theparagraph}{\Alph{paragraph}}
\newcommand{\bfa}{{\em A}}
\newcommand{\bfb}{{\em B}}
\newcommand{\bfas}{{\em A}'s~}
\newcommand{\bfbs}{{\em B}'s~}
\draft
\twocolumn[\hsize\textwidth\columnwidth\hsize\csname@twocolumnfalse%
\endcsname

\title{Interfacial mixing in heteroepitaxial growth\\}

\author{Boris Bierwald$^1$, Michael von den Driesch$^1$, Z\'eno
  Farkas$^1$, Sang Bub Lee$^{1,2}$, Dietrich E. Wolf$^1$\\[2mm]} 

\address{
  $1$ Institut f\"ur Physik, Universit\"at Duisburg-Essen, D-47048
  Duisburg, Germany\\ 
  $^2$ Department of Physics, Kyungpook National University, Taegu,
  702-701 Korea\\
 }
 
\maketitle
\begin{abstract}
  We investigate the growth of a film of some element {\em B} on a
  substrate made of another substance {\em A} in a 
  model of molecular beam epitaxy. A vertical
  exchange mechanism allows the {\em A}-atoms to stay on the growing
  surface with a certain probability. Using kinetic Monte Carlo 
  simulations as well as scaling arguments, the
  incorporation of the \bfas into the growing {\em B}-layer is
  investigated. Moreover we develop a rate equation theory for this process.
  In the limit of perfect layer-by-layer growth, 
  the density of {\em A}-atoms decays in the {\em B}-film like 
  (distance from the interface)$^{-2}$. The power law is cut off 
  exponentially at a characteristic thickness of the interdiffusion zone
  that depends on the rate of exchange of a {\em B}-adatom
  with an {\em A}-atom in the surface and on the system size. Kinetic 
  roughening changes the exponents. Then the thickness of the interdiffusion
  zone is determined by the diffusion length.
\end{abstract} \vspace{2mm}

]

\section{Introduction} 
 
Heterolayers, where e.g. ferromagnets are in contact with
antiferromagnets, semiconductors or superconductors,
give rise to new ordering and transport phenomena, which depend
crucially on the interfacial structure. Examples for such ordering
phenomena are the exchange bias \cite{Nowak02}, or the cryptomagnetism
\cite{Bergeret00}. Electronic transport
through a ferromagnetic-nonmagnetic-ferromagnetic sandwich (``spin valve''
geometry) gives rise to the giant magneto-resistance
\cite{Baibich88,Binasch89}. Another example is the 
recently predicted possibility to enhance or reduce a Josephson current
magnetically by replacing the tunnel barrier of a Josephson junction by a 
ferromagnetic-insulating-ferromagnetic sandwich \cite{Bergeret01}. These
phenomena belong to the growing field of spintronics
\cite{spintronics}, where the spin degree of freedom is used for  
electronic signal processing. Interfacial mixing affects all of them
\cite{Uzdin02}. For example, it leads to 
spin scattering disturbing the spin dependent transport properties. 

Therefore it is important to be able to control the various physical processes
that may spoil well defined interfaces. Some of them proceed after growth such 
as bulk interdiffusion or chemical interface reactions like silicide formation.
However, there are also important processes taking place exclusively at the  
surface: For example the 
substrate may partially behave like a surfactant, when one grows a 
different material on it. It is this latter mechanism which we investigate in 
this paper. The questions we want to answer concern the asymptotic
concentration profile, the width of the interdiffusion zone and
possible correlations among the substrate impurities within the
growing layer.

Specifically we consider growing some material \bfb~on a substrate \bfa. 
Obviously the interfacial mixing requires that some substrate (\bfa-) atoms get
replaced by \bfb-atoms and ``float up'' on the surface until they
get incorporated into the growing film. There is ample
experimental evidence that such a behaviour occurs in very different systems
like Cr on Fe \cite{Venus96}, AlAs on GaAs \cite{Lorke94}, Nb on Fe
\cite{Wolf03} or Au on Fe \cite{Bischoff01}. This process depends on
several important parameters including the lattice mismatch and the
interaction between the different atoms including magnetic
contributions. In particular the 
explanation of any ordering of \bfa- and \bfb-atoms close to the interface
would require a detailed investigation of these interactions
\cite{Schroeder02}. 

The situation becomes considerably simpler, however, if one is interested
in the physical properties further away from the interface.
Then the concentration of
\bfa-atoms may be regarded as sufficiently low that their interaction as
well as \bfa-\bfb-ordering become unimportant. The focus on this region
justifies our simplified model, in which the interdiffusion zone
depends only on the deposition rate $F$ of \bfb-atoms, the diffusion
constants $D_{\rm A}$ and $D_{\rm B}$ of the adatoms of type \bfa or \bfb, 
respectively, and the rate $E$ for the exchange of \bfb-adatoms with \bfa-atoms. 
The limit $E/D_{\rm B} \rightarrow \infty$,
where a \bfb-adatom exchanges with the first \bfa-atom it encounters, 
would be realized, if the \bfb-adatoms diffuse by an exchange
mechanism \cite{Kellog94},
while the \bfa-adatoms diffuse by hopping. 
In the present paper we simplify the model
even further by assuming that both kinds of atoms diffuse equally fast
on the surface, $D_{\rm A} = D_{\rm B} = D$, with a diffusion constant
independent of the surface composition.

Apart from the exchange there is a second crucial ingredient in the
model: The \bfa-atoms behave only {\em partially} as a surfactant in
the sense that they can be overgrown by  island edges. By contrast a
perfect surfactant atom should ``float up'' also in front of an
advancing island edge. 

Naively one would expect an exponential decay of the density profile
of \bfa-atoms far from the interface. It is the main result of this
investigation that this is not always the case: The incorporation of
\bfa-atoms is much slower, giving rise to a power law decay of the
concentration profile in the limit of perfect layer-by-layer growth,
$D/F \rightarrow \infty$. In this case the width of the interdiffusion
zone diverges, provided there are no finite size effects.  
By contrast, we shall show that for finite $D/F$ the width of the 
interdiffusion zone is no longer infinite, but a power law of $D/F$.

This paper is organized as follows. In the next section we are going
to define a simple solid-on-solid (SOS) model for epitaxial growth of
a \bfb-layer on an \bfa-substrate, which
allows for the irreversible exchange of \bfb-atoms with
\bfas at the surface. In this model the \bfa-atoms on the surface turn
out to cluster in a time-periodic self-organized way, which is
explained in Sec.\ref{sect:correl}. The next three sections,
Sec.\ref{mf1} -- \ref{numerics}, are devoted to the limit $D/F \rightarrow
\infty$. First, in Sec.\ref{mf1}, we present a simple mean field
argument leading to the prediction, that the 
concentration of \bfas decays algebraically 
in the \bfb-layer. For a finite system size this power-law is cut off
leading to a finite width $H$ of the interdiffusion zone, which is
discussed in Sec.\ref{width}, where also a scaling ansatz for the
surface concentration $c_A$ is proposed. This scaling ansatz is
confirmed by simulation results for one- and two-dimensional surfaces
in Sec.\ref{numerics}. 
The remaining sections deal with interdiffusion for finite $D/F$.
Sec.\ref{finite_dof} contains simulation results and scaling
arguments, and in Sec.\ref{rate_equ} a rate equation theory is
developed for the interdiffusion problem. In the Appendix we describe
a very efficient implementation of the simulation model for one-dimensional
surfaces in the limit $D/F \rightarrow \infty$.

\section{The model}
\label{sec:model}
In order to model heteroepitaxial growth of a \bfb-layer on an 
\bfa-substrate including the interfacial mixing, we introduce a simple
solid-on-solid (SOS) model, 
where the lattice mismatch and most interactions between the atoms 
are neglected. Moreover we describe the exchange mechanism of
\bfb-adatoms with \bfa-atoms in the surface simply by a
phenomenological constant exchange rate $E$, although it depends in
reality on the system parameters as well as the local environment.

\begin{figure}[ht]  
  \begin{center}
    \includegraphics[angle=0, width=75mm]{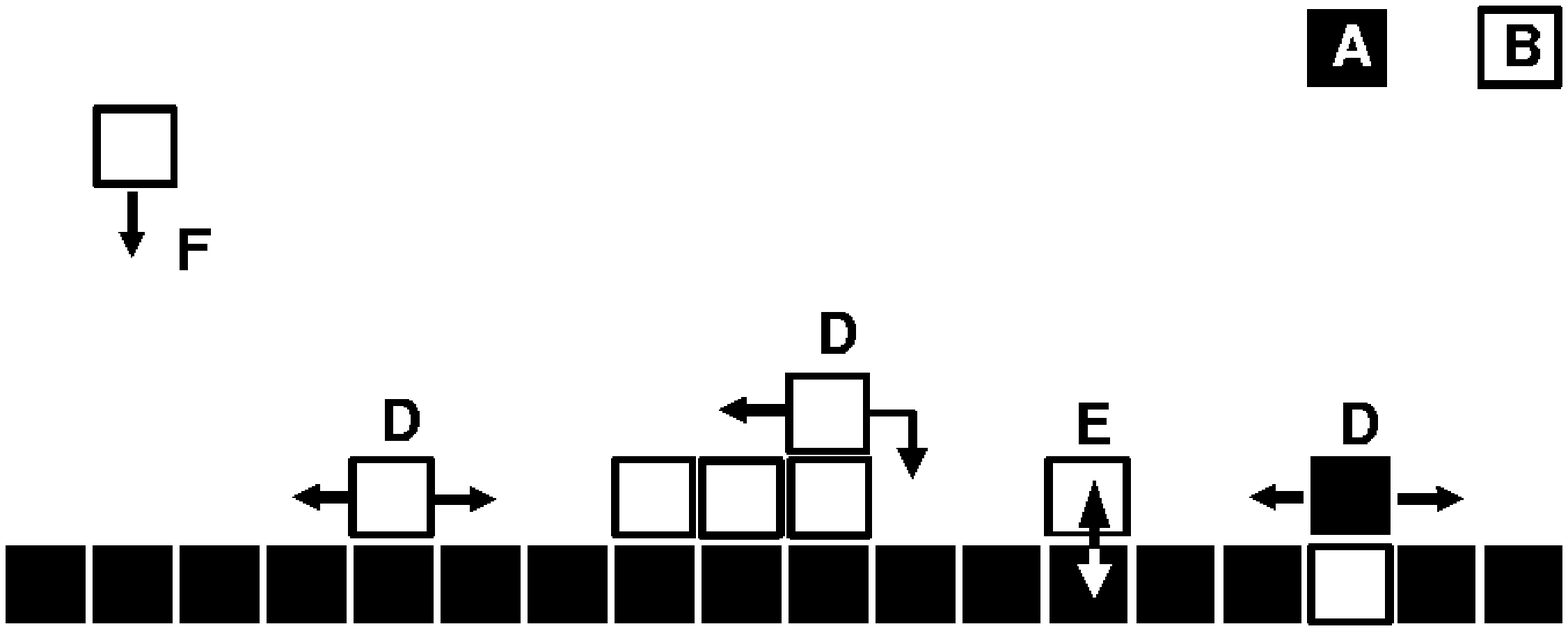}
    \vspace{2mm} 
    \smallcaption{
      \label{model}
      The growth process of the model. \bfb atoms are deposited with a
      deposition rate $F$ onto the substrate. Adatoms diffuse on the
      surface and down to the binding site at an island edge with
      diffusion constant $D$  irrespective of their type. In the top
      layer mobile \bfbs can exchange with \bfas with an effective exchange
      rate $E$.
      }
  \end{center}
\end{figure}

The model is defined on a simple cubic lattice by the following
kinetic rules (cf. Fig. \ref{model}):\\
(1) Starting from an initially flat substrate consisting of
\bfa-atoms, \bfb-atoms are deposited at randomly selected sites on the
surface with deposition rate $F$. \\
(2) As long as they do not have a lateral neighbor, the \bfb-atoms
diffuse on the surface with diffusion constant~$D$. \\
(3) When such a \bfb-atoms happens to sit on top of a \bfa-atom, it can
exchange vertically with rate $E$ or continue to diffuse with rate $D$.\\
(4)  After an exchange, the \bfb-atom stays irreversibly bound, whereas
the \bfa-atom diffuses on the surface with diffusion
constant $D$.\\
(5) There is {\em no} back exchange, when a \bfa-atom
sits on top of a \bfb-atom.\\
(6) When two adatoms, regardless of their type, meet, they 
form a stable non-moving nucleation center of an island.\\
(7) When an adatom, regardless of its type, reaches a site adjacent
to an island, it is irreversibly bound, increasing the size of the island.\\
(8) Both types of particles can diffuse down across terrace
edges without being hindered by an Ehrlich-Schwoebel barrier.\\
(9) There are no overhangs, i.e., we assume SOS growth.\\
(10) The exchange of a \bfb-atom with an \bfa-atom underneath is
forbidden, if the \bfb-atom is already part of an
island, i.e., if it has a nearest neighbor at the same
height. Hence \bfa-atoms can be overgrown by island edges.\\ 

Measuring time in units of monolayers (ML) and lengths in units of the lattice
constant $a$, this model is controlled by the two dimensionless parameters
$D/F a^4$ and $r_E \equiv a^2 E/D$. (In the following we set $a=1$.)  
 
This model will be investigated for one- and two-dimensional ($d=1$ and
$d=2$) surfaces in the following. Note that it reduces to the usual model
for MBE growth (for a recent review see e.g. 
\cite{Levi97}), if one does not distinguish the
two particle types. Therefore, the $D/F$-dependence of quantities like
island density, adatom density, surface width,  etc. are the same as usual. 
For general values of $D/F$ and $E/D$ we use kinetic Monte Carlo simulations 
\cite{Binder86,Newman99} in order to investigate the model. However,
for one-dimensional 
surfaces in the limit $D/F \rightarrow \infty$ we implemented a much more 
efficient algorithm, which is described in the Appendix.

\section{Correlations of the A-atoms}
\label{sect:correl}

\begin{figure}[ht]
  \begin{center}
    \includegraphics[angle=0, width=40mm]{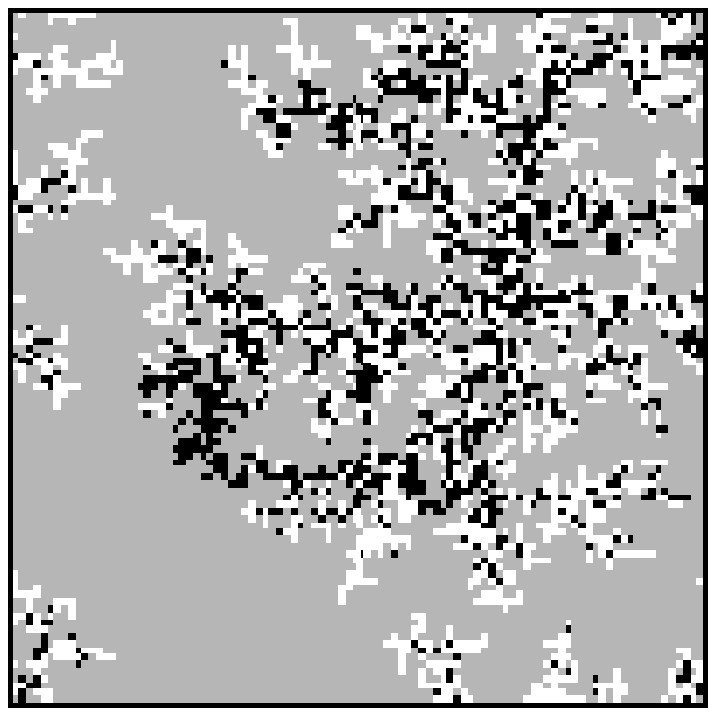}
    \includegraphics[angle=0, width=40mm]{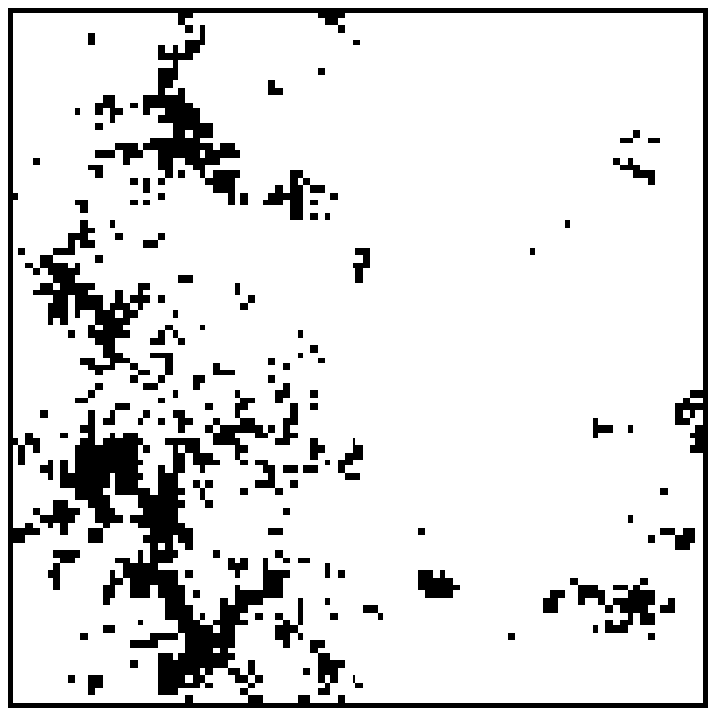}
    \vspace{2mm} 
    \smallcaption{
      \label{morph}
      Top view on the surface structure at $t=3.3$ ML (left) and 
      $t=4.0$ ML (right).
      $A$-atoms are black, $B$ atoms are height-encoded, where brighter 
      means higher.
      $D/F = 10^7 , E/D = 10^3 , L^2 = 100 \times 100$.
      }
  \end{center}
\end{figure}

\begin{figure}[ht]
  \begin{center}
    \includegraphics[angle=270, width=85mm]{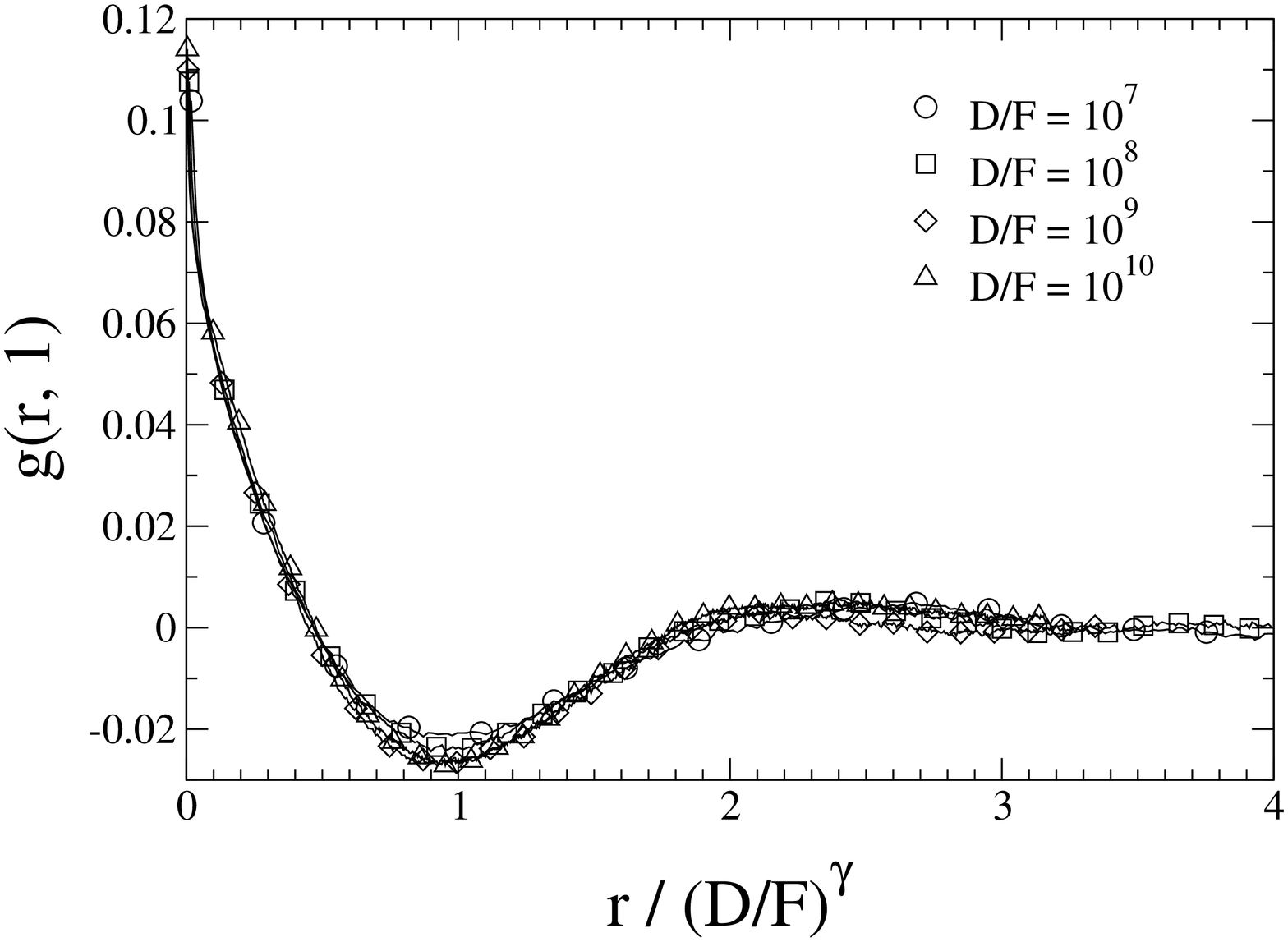}
    \vspace{2mm}
    \smallcaption{
      \label{correl}
      Rescaled lateral $A$-$A$ correlation functions $g(r,t)$ at $t=1$ 
      for different $D/F$ ($d=1$). The average distance between
      $A$-clusters scales with $D/F$ like the distance of the  
      nucleation sites, $\ell_{\rm D}$.
    }
  \end{center}
\end{figure}

One of the most intriguing qualitative properties of this model is 
the time-periodic self organization of \bfa-clusters on the growing surface 
with a period of one monolayer. Fig.\ref{morph} shows that the
\bfa-atoms (black) are first clustered around the nucleation sites
of a new layer, but migrate towards the holes remaining
in that layer, when the islands coalesce. Thus the
characteristic distance between these clusters agrees with the 
typical distance between the nucleation sites, the diffusion length
\begin{equation}
\ell_{\rm D}\sim (D/F)^{\gamma},
\label{eq:diffusion_length}
\end{equation}
as long as layer-by-layer growth persits.
This can be verified by examining the lateral correlations of $A$-atoms on the 
surface after deposition of $t$ monolayers:
\begin{equation}
  g(\vec r,t) = \frac{1}{L^d} \sum_{x=1}^{L^d} \rho_A(\vec x,t) \
  \rho_A(\vec x+\vec r,t)
  - c_A(t)^2 \ ,
\end{equation}
where $\rho_A(\vec x)$ denotes, whether there is an $A$-atom at the surface
at site $\vec x$ [$\rho_A(\vec x)$ = 1] or not [$\rho_A(\vec x)$ = 0],
and $c_A(t)$ denotes the surface density of $A$-atoms.  $d$ is the
dimension of the surface. In our simulations, $d$ was 1 or 2. A data
collapse of these correlation functions for different values of $D/F$
is obtained, if the space coordinates are rescaled by $\ell_{\rm D}$
(see Fig. \ref{correl} for $d=1$), which shows that this is the
characteristic distance between the \bfa-clusters.

The mechanism of the periodic self organization can most clearly be
seen in the limit 
$r_{\rm E}= E/D \rightarrow \infty$, where every $B$-adatom exchanges with
the first exchange partner $A$ it encounters.   
When layer $t$ is completed and layer $t+1$ begins to grow, the first
$B$-atoms deposited are likely to exchange with $A$-atoms from layer $t$.
Hence the nuclei of islands in the new layer $t+1$ will consist 
predominantly of $A$-atoms. As growth proceeds, the lower terrace (layer $t$)
gets depleted from exchange partners $A$, either because they are exchanged
with freshly deposited $B$-adatoms or get overgrown by the islands.
Then the core of the islands with a high concentration of $A$-atoms gets
surrounded by mainly $B$-atoms (Fig. \ref{morph}, left). 
However, as the island size increases, it becomes more and more likely
that $B$-atoms are deposited on top of the islands, 
i.e. in layer $t+2$. These $B$-adatoms find many exchange partners at the
core of the islands, which then become ``washed out'' and start decorating
the island edges, because there are no Ehrlich-Schwoebel barriers in our 
model. Note that this edge decoration with A-atoms happens {\em without
lateral exchange} of $B$- with $A$-atoms at the edges of the islands,
in contrast 
to the situation studied in \cite{Kotrla00}. As a result, the interior of the
islands gets cleared from $A$-atoms, which are collected in the holes of
layer $t+1$ which get filled last (Fig. \ref{morph}, right).
Then the process starts again: Layer 
$t+2$ nucleates predominantly with $A$-atoms which were exchanged from
layer $t+1$.  

For finite $r_{\rm E}$ the mechanism is similar. However, for vicinal
surfaces growing in step flow mode, the correlations among the $A$-atoms
are different. Here, the terraces get cleared of $A$-atoms, 
which attach to the step edges. 
This decoration of advancing edges leads to a correlation pattern
$g(\vec r,t)$ with a spatial periodicity  identical to the width of the
terraces. 

\section{The scale free limit} 
\label{mf1}

The limit of perfect layer-by-layer growth, $D/F \rightarrow \infty$,
is particularly instructive. In this case, there is only one island on
the surface, nucleating at a random site. Afterwards
at most one adatom can be found on the surface at a time. 
If we assume maximal exchange in addition, i.e. an exchange rate $E$
much larger than both $D$ and $F$, $r_E = E/D \rightarrow \infty$,
the model becomes parameter-free. \bfa-atoms can be buried even in this
limit, when they are overgrown by an island edge. However,
the last \bfa-atom will never get buried in this case. The nucleus of
a new layer will always contain at least one of the 
remaining surface atoms of type \bfa in the limit we focus on, because
the first \bfb-atom deposited after completion of a layer exchanges with
an \bfa-atom before the next \bfb-atom gets deposited: $c_A(t
\rightarrow \infty) = L^{-d}$.  As this case is
scale free, we expect that the concentration of \bfa-atoms falls off
like a power law into the growing \bfb-film. 

In order to get a first idea about the distribution of {\em A}-atoms in the
growing {\em B}-film, it suffices to consider the concentration $c_A(n)$
of $A$-atoms at the surface after the deposition of $n$
monolayers. During the growth of the next layer, a certain fraction $(1-q)
\in[0,1)$ of these $A$-atoms is transported to the next layer via
vertical exchange:
\begin{equation}
  \label{mf-ansatz}
  c_A(n+1)=(1-q)c_A(n).
\end{equation}
If $q$ was constant, this would imply an exponential decay
$c_A(n) \propto (1-q)^n$.  In the present model, however, the
probability $q$ that an $A$-atom gets overgrown decreases with
decreasing $c_A(n)$, resulting in a decay which is slower than
exponential. 
 
Qualitatively this can be understood in the following way: In the
limit $D/F \rightarrow \infty$ there is only one island on the 
surface. Moreover, any $A$-atom on the lower terrace gets transferred to
the new layer as soon as it is reached by an adatom of type $B$,
because $r_{\rm E} \rightarrow \infty$ is assumed, as well. Only
$A$-atoms sufficiently close to the island have a chance to be overgrown
by the island edge before being visited by a $B$-atom. 

The island edge advances a characteristic distance $\ell_{\rm cover}$,
while the lower terrace gets depleted from $A$-atoms by exchange with
adatoms of type $B$. For a one-dimensional surface of length $L$ it is clear that
$\ell_{\rm cover}$ is proportional to the number of $A$-atoms at the
surface, $c_A(n)L$. Hence, the number of $A$-atoms with a chance to be
overgrown is of the order of
\begin{equation}
\ell_{\rm cover} c_A(n) \propto c_A(n)^2 L.
\end{equation}
This must be compared with $q c_A(n) L$, which shows that
\begin{equation}
q \propto c_A(n).
\label{eq:q}
\end{equation}

Inserting Eq.(\ref{eq:q}) into Eq.(\ref{mf-ansatz}) leads to the difference equation
\begin{equation}
c_A(n+1)-c_A(n)\propto -c_A(n)^{2}
\end{equation}
implying the asymptotic power law 
\begin{equation}
  c_A(n)\sim 1/n 
\label{eq:exp-1}
\end{equation} 
for the concentration of $A$-atoms at the surface.

The concentration profile of $A$-atoms inside the grown film is given by
\begin{equation}
\rho(n) = c_A(n) - c_A(n+1) \sim 1/n^2.
\end{equation}
It is remarkable that for the mechanism discussed in this paper the
width of the interdiffusion zone diverges logarithmically with the
thickness  $T$ of the film:
\begin{equation}
\sum_{n=1}^{T-1}\rho(n)n + c_A(T) T \sim \ln{T}.
\end{equation}
Nevertheless, the interface can be localized precisely, because
$B$-atoms do not occur below layer $n=0$ due to the absence of bulk
diffusion in this model. Below we show that these power laws are
confirmed by simulations. 

The argument leading to Eq.(\ref{eq:q}) ignores that the $A$-atoms at the
surface are clustered, as shown in Fig.\ref{morph}, and was made
plausible only for a one-dimensional surface. However, it can be
refined such that it takes these spacial correlations into account and
applies also for two-dimensional surfaces. For $D/F \rightarrow \infty$, we can
imagine that there is
only one cluster of size $c_A(n) L^d$ on the surface, when the new
layer nucleates. The important point is that the nucleation happens 
anywhere on the surface with equal probability $1/L^d$ in this case. 
However, only if the nucleation site is within an area of about the
size $2^d
c_A(n) L^d$ centered at the middle of the cluster, there is a chance
that some $A$-atoms get overgrown. In other words, only a fraction of
nucleation sites $\propto c_A(n)$ leads to overgrowth. The average
number of $A$-atoms overgrown in such a case is proportional to the
cluster size. Hence on average a fraction $q$ of $A$-atoms is
overgrown which is
proportional to the fraction of nucleation sites leading to
overgrowth, i.e. this refined argument gives $q \propto c_A(n)$ 
as in Eq.(\ref{eq:q}).

\section{The width of the interdiffusion zone for $D/F
  \rightarrow \infty$} 
\label{width}

In the previous section we predicted that the width of the
interdiffusion zone diverges in the scale free limit, where
$D/F \rightarrow \infty$ and $r_{\rm E} \rightarrow \infty$.
In this section we predict, that for finite $r_{\rm E}$ 
the power law Eq.(\ref{eq:exp-1}) is only valid, if the system is
infinitely large. For finite system size the power law is 
exponentially cut off at a characteristic
width of the interdiffusion zone.

After the completion of several monolayers on a substrate of linear size $L$ 
the number of substrate atoms at the surface is $c_A L^d$.  The $A$-atoms
are concentrated in a cluster, which we assume to  be compact, hence of
diameter $\propto c_A^{1/d} L$. This assumption is justified even for $d=2$, 
where the islands initially are fractal, because the $A$-cluster occupies the 
sites which were filled {\em last} in the uppermost monolayer. These sites do 
not form a fractal.

Now we imagine the surface to be coarse grained on the scale of the cluster
diameter so that exactly one cell contains the $A$-cluster. The
typical residence time
of an adatom in such a cell is 
\begin{equation}
\Delta t = \frac{(c_A^{1/d} L)^2}{D}.
\end{equation} 

A $B$-adatom which enters the cell containing the $A$-cluster will almost 
certainly be replaced by an exchange partner $A$ within the residence time,
if
\begin{equation}
\label{width:condition}
E\Delta t = r_E (c_A^{1/d} L)^2 \gg 1
\end{equation}
(exchange dominates).
For $E\Delta t \ll 1$ the adatom changes from type $B$ into type $A$ only with
probability $E\Delta t$ (overgrowth dominates).

It is plausible to assume that the power law belongs to the exchange dominated
slow decay of $c_A$ while the exponential cut off indicates the much faster
decay when overgrowth dominates. Thus the width $H$ of the interdiffusion zone
should be reached, when $c_A $ becomes so small that exchange is no longer
guaranteed, i.e. when $E\Delta t$ drops below 1. Inserting 
$c_A \approx 1/t = 1/H$ into Eq.(\ref{width:condition}) one obtains
\begin{equation}
\label{width:h}
H \approx (\sqrt{r_E} L)^d.
\end{equation}
However, as $c_A$ cannot become smaller than $L^{-d}$, this estimate is only
valid for $r_E < 1$, while $H \approx L^d$ for $r_E > 1$.
This is our prediction for the width of the interdiffusion zone in the limit
$D/F \rightarrow \infty$. Note that in this limit the cutoff of the power law
is a finite size effect: For $L \rightarrow \infty$ the power law extends to
infinity. 

Based on the results of this and the previous paragraph we can
conjecture the following scaling form for the surface concentration of
\bfas:
\begin{equation}
\label{width:scaling}
c_A(t,L;r_E) - c_A(t\rightarrow \infty) = \frac{1}{H}
f\left( \frac{t}{H} \right) 
\end{equation}
where according to Eq.(\ref{eq:exp-1})
\begin{equation}
\label{width:f}
f(\tau) \sim 1/\tau \quad {\rm for} \quad \tau \ll 1.
\end{equation}

\begin{figure}[ht]
  \begin{center}
    \includegraphics[angle=270, width=85mm]{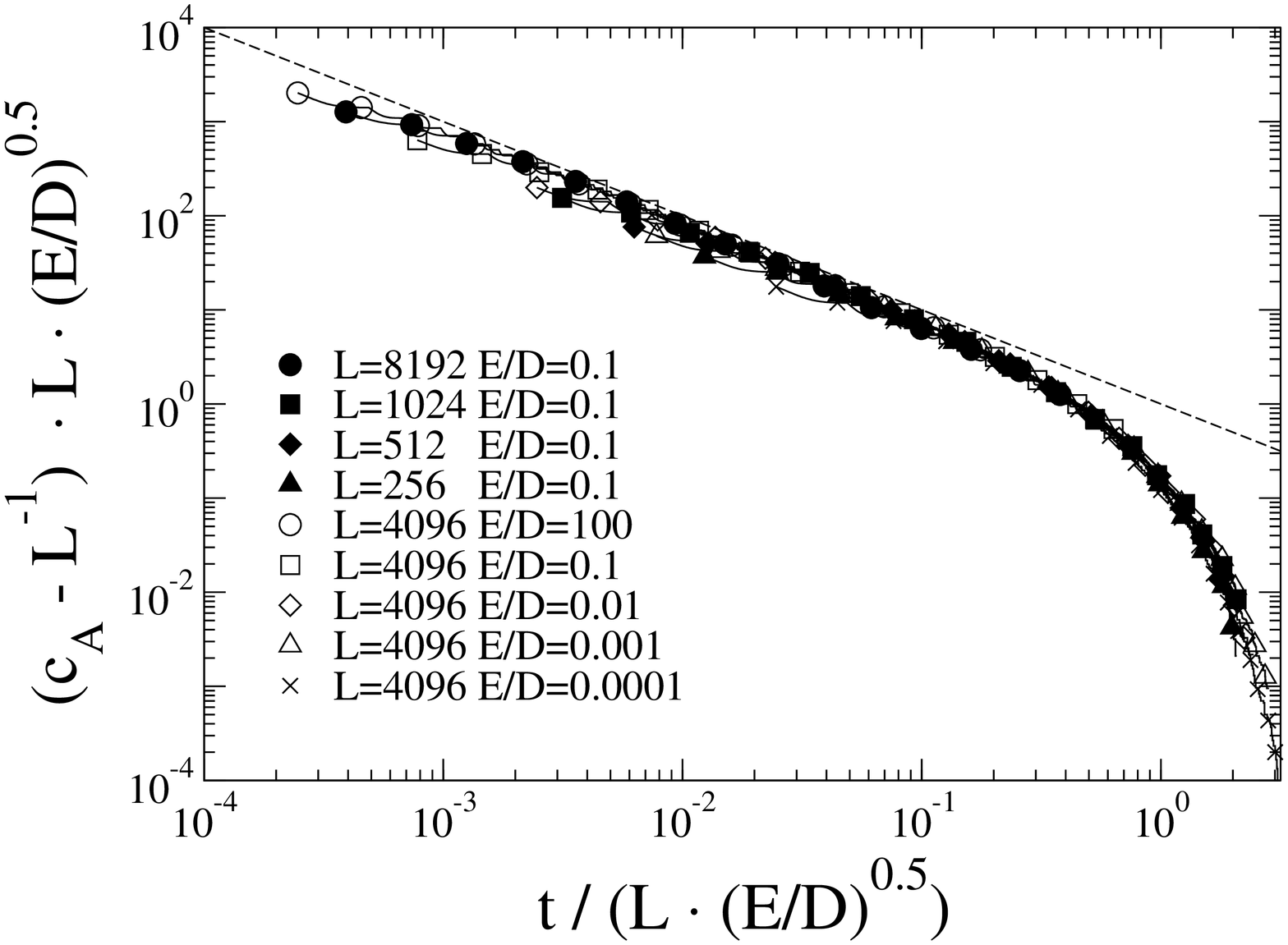}
    \vspace{2mm}
    \smallcaption{
      \label{scaling_inf_1d}
      Scaling of $c_A(t)$ with $L$ and $E/D$ in $d=1$ 
      for $D/F \to \infty$. Full
      symbols mark four curves for $E/D = 0.1$ and $L$ between 256 and
      8192. Open symbols mark five curves for $L=4096$ and $E/D$
      between $10^{-4}$ and 100. A data collapse of all 9 curves is
      reached by scaling in accordance with Eq.(\ref{width:scaling}) and
      Eq.(\ref{width:h}). The data for $E/D = 100$ were rescaled
      differently: Here, $c_A L -1$ is plotted vs. $t/L$, as if
      $E/D$ was 1 instead of 100. This shows, that $H$ becomes
      independent of $E/D$ for $E/D > 1$ as explained after Eq.(\ref{width:h}).
      The dashed line has slope -1 in agreement with Eq.(\ref{width:f}). 
    }
  \end{center}
\end{figure}

\section{Numerical Results for $D/F \rightarrow \infty$} 
\label{numerics}

In order to check the predictions of Secs. \ref{mf1} and \ref{width},
we simulated the model described in Sec.\ref{sec:model} for
$D/F\rightarrow\infty$ and varied the values of $r_{\rm E}$ and 
system size $L$ for one- and two-dimensional surfaces. For the case $d=1$
we used the algorithm described in the appendix, while kinetic Monte-Carlo
simulations were done for $d=2$. 

\begin{figure}[ht]
  \begin{center}
    \includegraphics[angle=270, width=85mm]{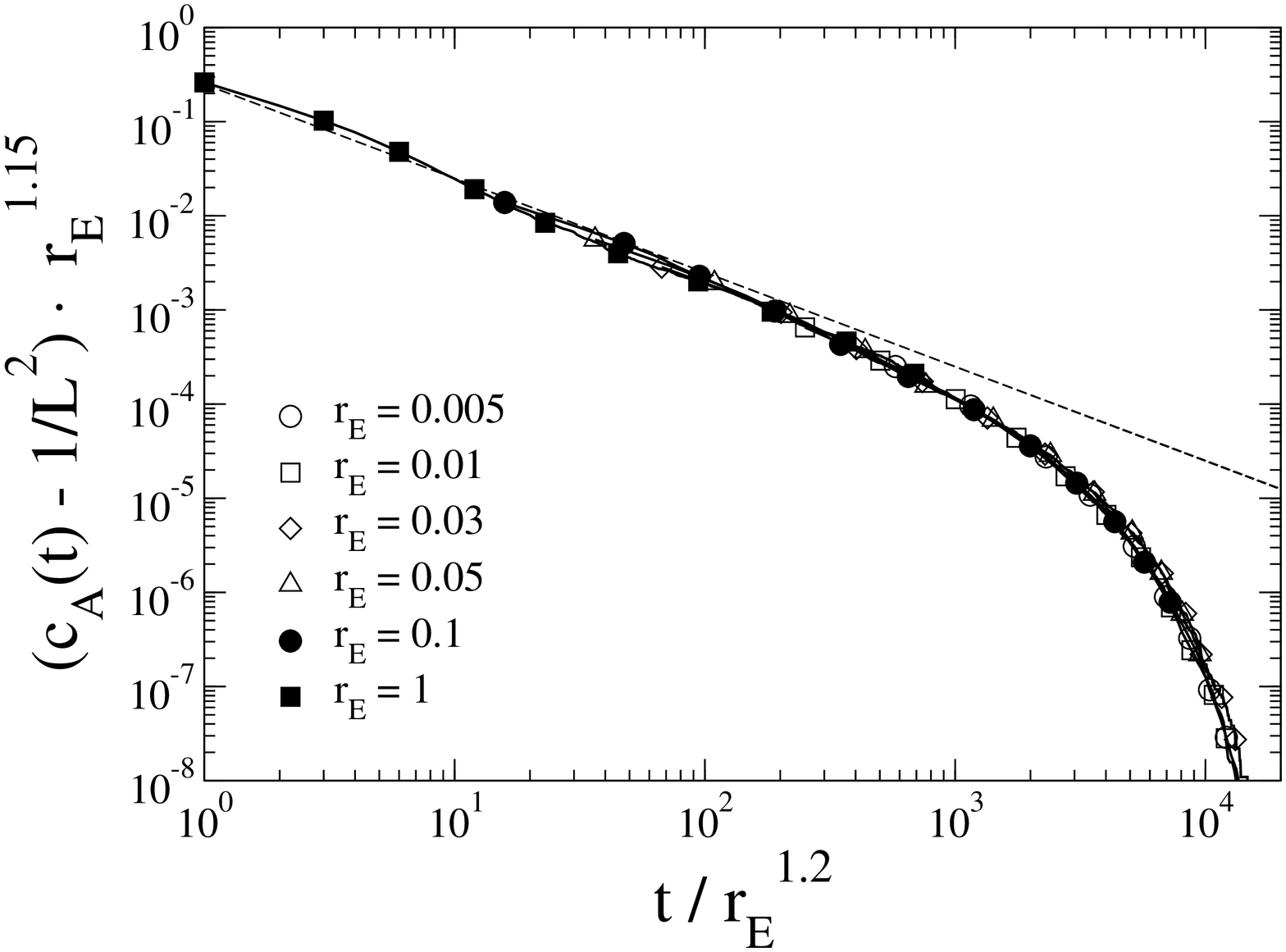}
    \vspace{2mm}
    \smallcaption{
      \label{scaling_inf_2d_eod}
      Scaling of $c_A(t)$ with $r_E=E/D$ in $d=2$ for $D/F \to \infty$.
      $L^2 = 200 \times 200$.
    }
  \end{center}
\end{figure}

\begin{figure}[ht]
  \begin{center}
    \includegraphics[angle=270, width=85mm]{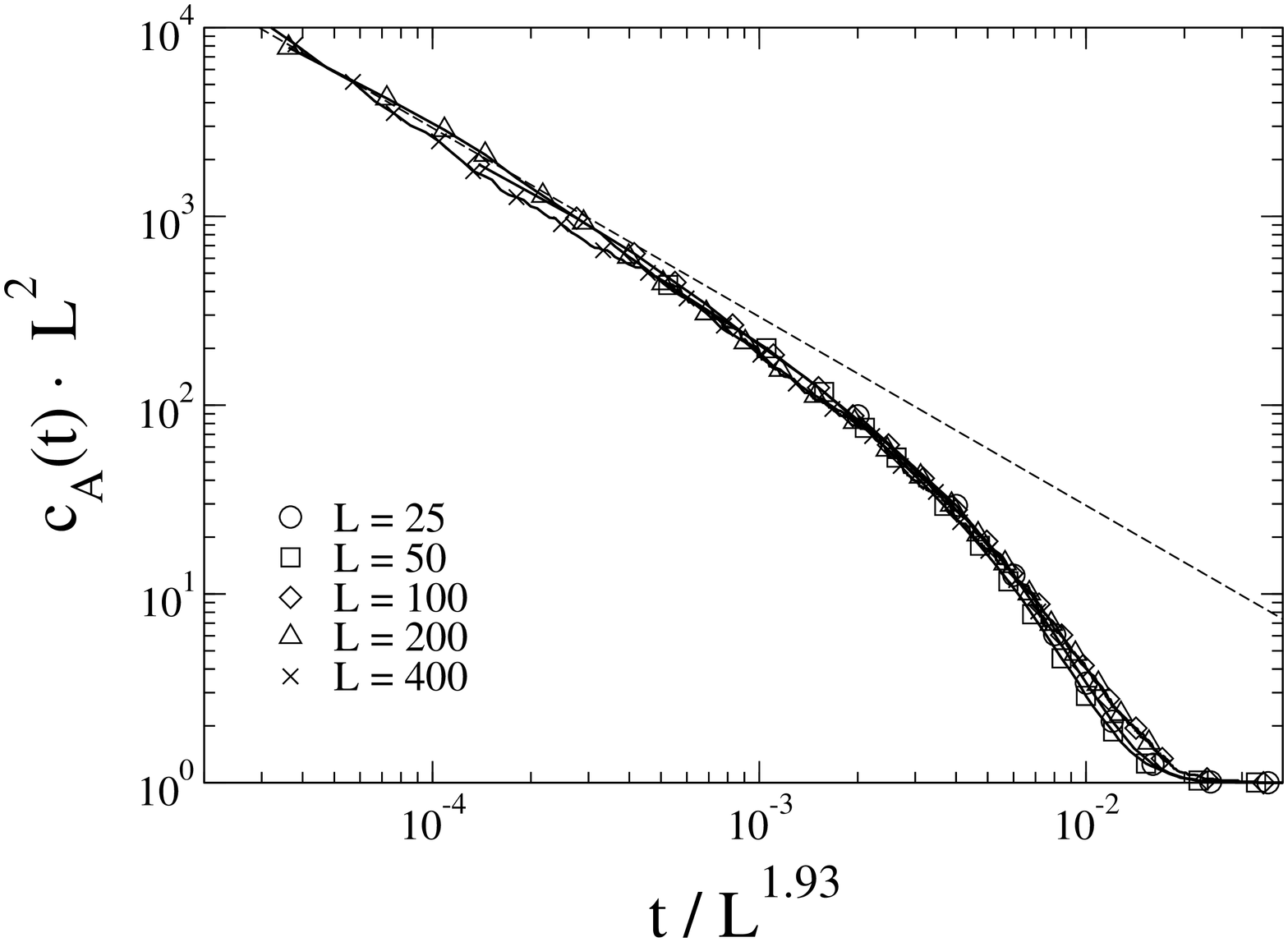}
    \vspace{2mm}
    \smallcaption{
      \label{scaling_inf_2d_l}
      Scaling of $c_A(t)$ with $L$ in $d=2$ for $D/F \to \infty$.
      $E/D = 0.1$.
    }
  \end{center}
\end{figure}

Fig. \ref{scaling_inf_1d} (for $d=1$) and 
Figures \ref{scaling_inf_2d_eod}, \ref{scaling_inf_2d_l} (for $d=2$) 
show the concentration $c_A$ of \bfa-atoms at the surface as a funcion
of deposition time $t$ (in monolayers of \bfb-atoms). All curves are 
averages over 200 - 400 independent runs. Both for $d=1$ and $d=2$ 
the exponent of the power law decay was found to be consistent with 
the value -1 derived in section \ref{mf1}.

As shown in Fig. \ref{scaling_inf_1d}, the predicted relations
Eq.(\ref{width:h}) and Eq.(\ref{width:scaling}) lead to the expected data
collapse for the one dimensional surface. 
The results in two dimensions are not as clear.  In this case,
we obtain the best data collapse with
\begin{equation}
  H \propto L^{1.93} \cdot r_{\rm E}^{1.2} \ ,
\end{equation}
as shown in the Figures \ref{scaling_inf_2d_eod} and \ref{scaling_inf_2d_l},
whereas our predicted exponents (2 and 1, respectively, see
Eq.(\ref{width:h})) were about 4\% and 20\% different. 
In fact the $A$-clusters are not as compact as assumed in the simple
argument of Sec.\ref{width} (see Fig.\ref{morph}).

\section{Scaling for finite $D/F$}
\label{finite_dof}

For finite $D/F$ there are many $A$-clusters on the surface. Their
typical distance is given by the diffusion length $\ell_{\rm D}$ as
shown in Sec.\ref{sect:correl}.  The average size of the $A$-clusters is
$c_A \ell_D^d$, provided this is much larger than 1. If the
concentration $c_A$ becomes too small, less and less $A$-clusters 
will be found on the surface, and their typical distance will
grow. Finally all $A$-atoms will be overgrown, 
in contrast to the situation of perfect layer-by-layer growth, where
the last $A$-atom could never be overgrown. Apart from this, 
one might expect that the results of the previous three
sections would essentially remain true, if one replaces $L$ by
$\ell_{\rm D}$. Qualitatively, the surface concentration
$c_A$ indeed decays first approximately as a power law of the
deposition time, which is cut off at a characteristic width $H$ of the
interdiffusion zone. 

\begin{figure}[ht]
  \begin{center}
    \includegraphics[angle=270, width=85mm]{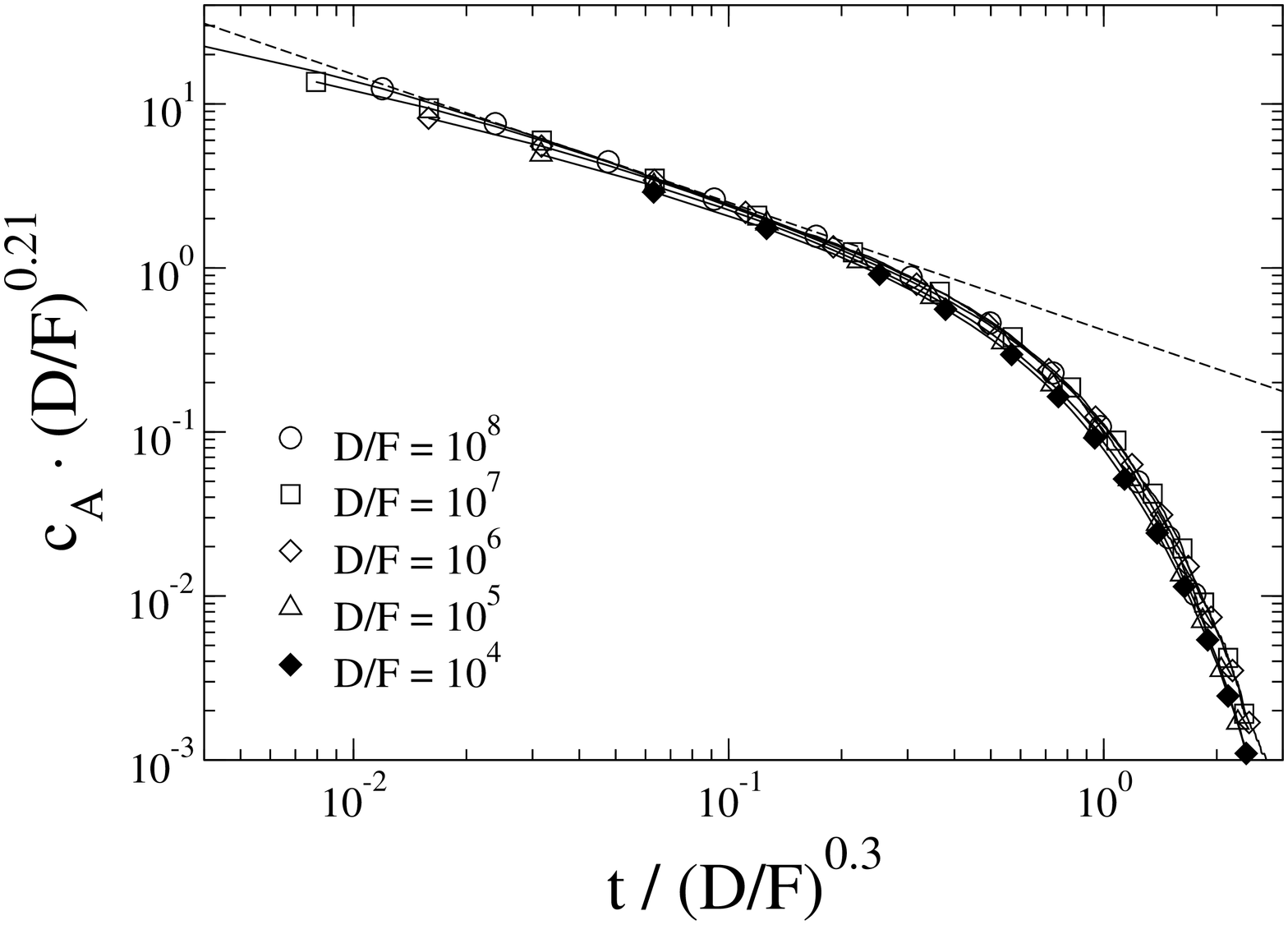}
    \vspace{2mm}
    \smallcaption{
      \label{finite_1d_dof}
      Scaling of $c_A(t)$ with $D/F$ in $d=1$.
      $E/D = 10^3 , L = 5 \cdot 10^3 \dots 10^4$. The dashed line
      indicates the exponent $\beta = 0.78$.
    }
  \end{center}
\end{figure}

\begin{figure}[ht]
  \begin{center}
    \includegraphics[angle=270, width=85mm]{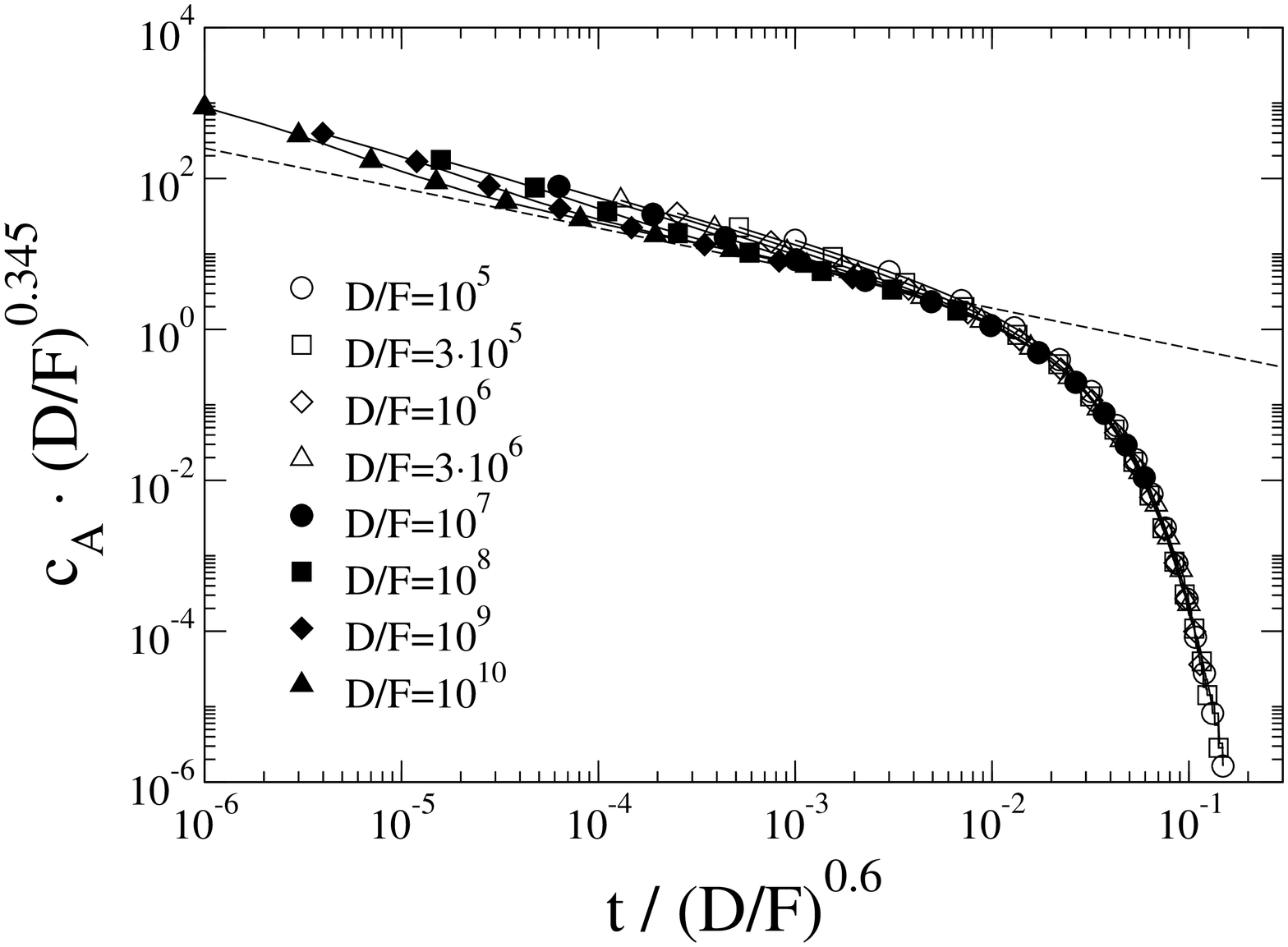}
    \vspace{2mm}
    \smallcaption{
      \label{finite_2d_dof}
      Scaling of $c_A(t)$ with $D/F$ in $d=2$.
      $E/D = 10^4 , L^2 = 500 \times 500$. The dashed line indicates
      the exponent $\beta = 0.53$.
    }
  \end{center}
\end{figure}

\begin{figure}[ht]
  \begin{center}
    \includegraphics[angle=270, width=85mm]{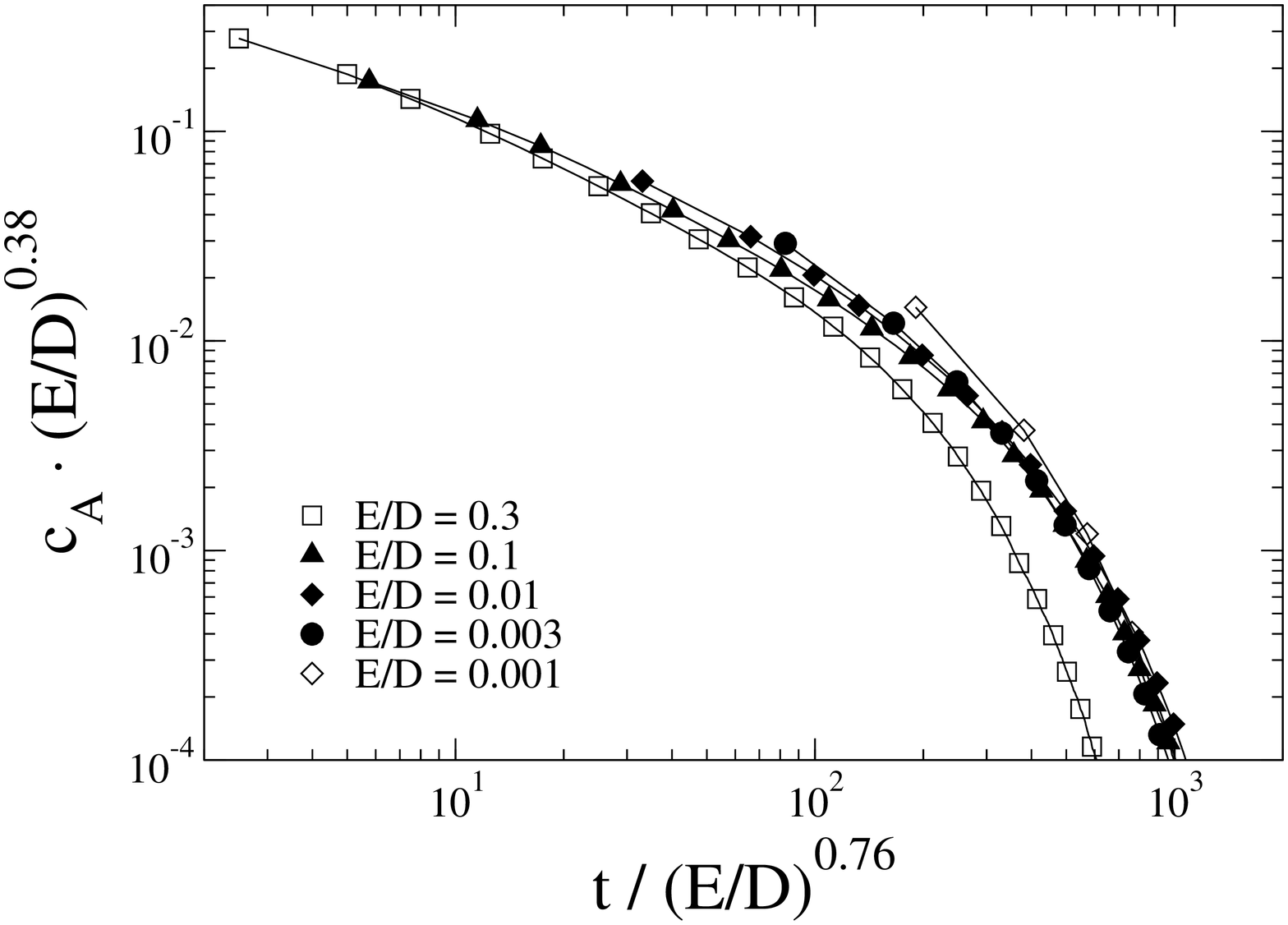}
    \vspace{2mm}
    \smallcaption{
      \label{finite_1d_eod}
      Scaling of $c_A(t)$ with $E/D$ in $d=1$.
      $D/F = 10^7 , L = 5 \cdot 10^3$.
    }
  \end{center}
\end{figure}

\begin{figure}[ht]
  \begin{center}
    \includegraphics[angle=270, width=85mm]{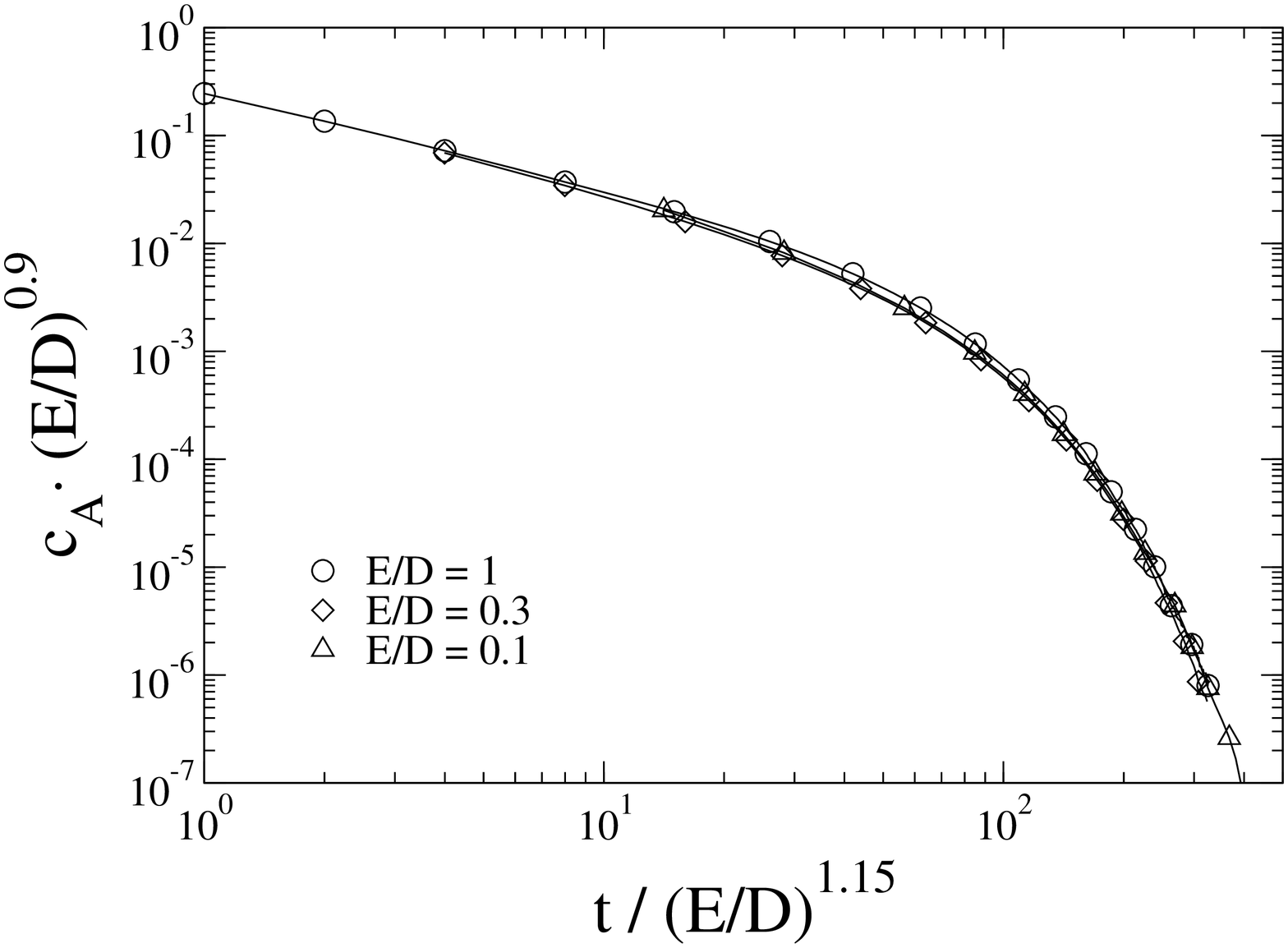}
    \vspace{2mm} 
    \smallcaption{
      \label{finite_2d_eod}
      Scaling of $c_A(t)$ with $E/D$ in $d=2$.
      $D/F = 10^7 , L^2 = 500 \times 500$.
    }
  \end{center}
\end{figure}

Quantitatively, however, the situation turns out to be more complex
than this: All exponents are different, as the simulation results
(Figs. \ref{finite_1d_dof} -- \ref{finite_2d_eod}) show.
The power law decay of $c_A \propto t^{-\beta}$ extends over at
most two decades for the largest values of $D/F$ we simulated, so that
the determination of the exponent $\beta$ from the slopes in the
log-log-plots Fig.\ref{finite_1d_dof} and \ref{finite_2d_dof} is not
very accurate. We estimate 
\begin{equation}
\beta= \left\{\begin{array}{lll}
              0.78 \pm 0.08 & {\rm for}& d=1,\\
              0.53 \pm 0.05 & {\rm for}& d=2,\\
              \end{array}\right.
\end{equation}
which are indicated by the dashed lines
in the two Figures. Both exponents are significantly smaller than 
$\beta = 1$ obtained for infinite $D/F$, i.e. $c_A$ decays more
slowly for finite than for infinite $D/F$.

This result is surprising on first sight, because there are more
island edges on the surface for finite $D/F$, hence more possible
places, where $A$-atoms may be overgrown. That $c_A$ decays more
slowly nevertheless, may be explained by the fact, that
the nucleation of islands does not happen anywhere with equal
probability as for infinite $D/F$ , but preferentially far away from
the holes in the previous layer, where the
{\em A}-atoms are concentrated.  Therefore overgrowth is less likely, and
$c_A$ decays more slowly than for infinite $D/F$. 

This raises the question, how big the parameter $D/F$ must be in order
to see the exponent $\beta = 1$ instead of the smaller one. The answer
is, that the system size $L$ must be small compared to $\ell_{\rm D}$
in order to obtain the crossover to the faster decay of $c_A$. 
This was confirmed by simulation results in \cite{Bierwald02}. 
In the four Figures belonging to this Section we carefully checked
that the system sizes were big enough to exclude finite size effects.

In analogy to Eq.(\ref{width:condition}) we expect that the power law
decay of $c_A \propto t^{-\beta}$ stops, when
\begin{equation}
r_{\rm E}(c_A \ell_{\rm D}^d)^{2/d} \approx 1, 
\end{equation}
or 
\begin{equation}
c_A \ell_{\rm D}^d \approx 1,
\end{equation}
whichever happens first. Replacing $c_A$ by $H^{-\beta}$, this 
implies that the width $H$ of the interdiffusion zone should be given
by 
\begin{equation}
H \approx \ell_{\rm D}^{d/\beta} \quad {\rm for } \quad r_{\rm E} \gg
1 
\label{eq:H1}
\end{equation}
and
\begin{equation}
H \approx (\sqrt{r_{\rm E}}\ell_{\rm D})^{d/\beta} \quad {\rm for } 
\quad r_{\rm E} \ll 1 .
\label{eq:H2}
\end{equation}
In analogy to Eq.(\ref{width:scaling}) we postulate then that
\begin{equation}
c_A = \frac{1}{H^{\beta}} g\left( \frac{t}{H} \right) 
\label{refined_scaling}
\end{equation}
with 
\begin{equation}
g(\tau) \sim 1/\tau^{\beta} \quad {\rm for} \quad \tau \ll 1,
\end{equation}
because $c_A$ is independent of $H$ for small $t$.

We first checked these conjectures for $r_{\rm E} > 1$, where the
surface concentration of \bfas becomes independent of $r_{\rm E}$, as
expected. If we insert the $D/F$-dependence
Eq.(\ref{eq:diffusion_length}) of the diffusion length in Eq.(\ref{eq:H1}),
Eq.(\ref{refined_scaling}) can be written in the form
\begin{equation}
c_A \left( \frac{D}{F} \right)^{\gamma d} = g_1
\left(t\left(\frac{D}{F} \right)^{-\gamma d/\beta}\right).
\end{equation}
With $\gamma=1/4$ \cite{Pimpinelli92} for $d=1$
and $\gamma = 1/(4+d_{\rm f}) \approx 0.17$ for $d=2$ ($d_{\rm f}$ is the
fractal dimension of the islands) \cite{Villain92}, and with the
$\beta$-values determined above, the theory predicts
\begin{equation}
\gamma d = \left\{\begin{array}{lll}
                 0.25          & {\rm for} & d=1\\
                 0.35 \pm 0.01 & {\rm for} & d=2\\
                 \end{array}\right.
\end{equation}
\begin{equation}
\gamma d/\beta = \left\{\begin{array}{lll}
                 0.32 \pm 0.03 & {\rm for} & d=1\\
                 0.66 \pm 0.07 & {\rm for} & d=2\\
                 \end{array}\right.
\end{equation}
The data collapses in Figs.\ref{finite_1d_dof}, \ref{finite_2d_dof}
are in reasonable agreement with this prediction.

However, the $r_{\rm E}$-dependence Eq.(\ref{eq:H2}) for $r_{\rm E} < 1$ 
is not in agreement with the simulation results. For fixed $D/F$ 
the theory Eq.(\ref{refined_scaling}) predicts
\begin{equation}
c_A r_{\rm E}^{d/2}= g_2(t \, r_{\rm E}^{-d/2\beta}) \, .
\label{r_E-scaling-theory}
\end{equation}
Inserting the value of $\beta$ determined above, the scaling exponent
should be
\begin{equation}
d/2\beta = \left\{\begin{array}{lll}
                 0.64 \pm 0.06 & {\rm for} & d=1\\
                 1.9 \pm 0.2 & {\rm for} & d=2\\
                 \end{array}\right.
\end{equation}
For $D/F = 10^7$ we could only check this for about one decade of 
$r_{\rm E}$-values: For $d=1$ we found that already for $r_{\rm E}=0.3$
the crossover into the regime, where $H$ becomes independent of
$r_{\rm E}$, affects the data. For $r_{\rm E}< 10^{-3}$ the exchange
was so weak that the surface concentration of \bfas decayed
very fast from the beginning, so that a convincing data collapse
was not possible. Similar problems occurred for d=2. The best result
of our attempts to get a data collapse in the available $r_{\rm
  E}$-interval are shown in Fig.\ref{finite_1d_eod} for $d=1$ and 
Fig. \ref{finite_2d_eod} for $d=2$. The effective exponents turn out
to be very different from the ones predicted in
Eq.(\ref{r_E-scaling-theory}). 

\section{Rate Equation Approach}
\label{rate_equ}

In this Section we extend the established rate equation approach for
submonolayer homoepitaxial growth as described in \cite{Tang93,Amar94},
in order to apply it to our model for surface interdiffusion. 
Our approach describes the time evolution of 
four submonolayer quantities:  
The density of mobile adatoms $\rho$, the total island density $I$, 
the density of mobile adatoms of type $B$, 
$\rho_B$, and the density of potential exchange partners of type $A$ in
the lower layer, $\rho_A$.
With these quantities, the rate equations are as follows:
 
\begin{eqnarray}
  \label{eq:dgl-sys-1}
  \dot \rho &=& F - D \rho ( I + 2 \rho ),\\
  \label{eq:dgl-sys-2}
  \dot I    &=& D \rho^2,\\
  \label{eq:dgl-sys-3}
  \dot \rho _B &=& F - D \rho_B ( I + \rho + \rho_B) 
  - E a^2 \rho_A \rho_B,\\
  \label{eq:dgl-sys-4}
  \dot \rho _A &=& -E a^2 \rho_A \rho_B - 
  \rho_A a^2 D \rho ( I + 2 \rho ).
\end{eqnarray}

While the first two equations are identical to the well known point-island 
model rate equations for homoepitaxial growth,
the last two are specific to our heteroepitaxial model.
The third one expresses the change in density of the $B$-type adatoms.
Its positive contribution describes the depostition of new adatoms. 
The first negative term represents the loss of {\em B}-adatoms, when
they get incorporated into islands or bind to another adatom to
nucleate a new island. The extra term $D\rho_B^2$ accounts for the
fact that nucleation events involving two {\em B}-adatoms count twice
as much as those between a {\em B}- and an {\em A}-adatom, because
they remove two {\em B}-adatoms simultaneously.
The last term of Eq.(\ref{eq:dgl-sys-3})
describes the exchange of mobile $B$'s with $A$'s.
Equation (\ref{eq:dgl-sys-4}) expresses the annihilation of possible $A$-type  
exchange partners 
in the lower layer. Since this value is monotoneously decreasing,
there is no  
positive contribution. The negative terms describe the exchange of $A$'s
with mobile $B$'s, and the overgrowth of $A$'s due to propagating island 
edges and nucleation events.

Rescaling the variables (see \cite{Tang93}) according to 
$\hat t      = t F  \ell_0^2$, 
$\hat \rho   = \rho \ell_0^2$,
$\hat I      = I    \ell_0^2$,
$\hat \rho_A = \rho_A \ell_0^2$, 
$\hat \rho_B = \rho_B \ell_0^2$,
where $\ell_0 = (D/F)^{1/4}$, leads to the dimensionless equations
\begin{eqnarray}
\label{eq:dgl-sys-scale-1}
\dot{\hat{\rho}} &=& 1 - \hat \rho ( \hat I + 2 \hat \rho ),\\
\label{eq:dgl-sys-scale-2}
\dot{\hat{I}}    &=& \hat{\rho}^2,\\
\label{eq:dgl-sys-scale-3}
\dot{\hat{\rho}}_B &=& 1 - \hat\rho_B ( \hat I + \hat \rho + \hat \rho_B) 
  - r_E \hat \rho_A \hat \rho_B,\\
\label{eq:dgl-sys-scale-4}
\dot{\hat{\rho}}_A &=& - r_E \hat \rho_A \hat \rho_B
  - \left(a/l_0\right)^2 \hat \rho_A \hat \rho ( \hat I + 2
  \hat \rho ).
\end{eqnarray} 

If we consider systems in perfect layer-by-layer growth mode, this 
approach not only holds for the submonolayer regime starting from the 
substrate, but also starting after integer numbers of deposited monolayers.
The initial conditions of these equations for a flat surface
after $n$ deposited monolayers are
\begin{equation}
  \hat{\rho}(0) = \hat{I}(0) = \hat{\rho}_B(0)= 0 , \qquad
  \hat{\rho}_A(0) = \hat{c}_A(n) \, . 
\end{equation}

Disregarding the point island model nature of this approach,
which only holds for early stages of the submonolayer regime, 
we can establish the surface concentration of \bfas after the deposition of
one additional monolayer, $c_A(n+1)$, as the integral over the density 
of all exchanged atoms:
\begin{equation}
  \label{eq:rek-c_A}
  \hat{c}_A(n + 1) = \int_0^{(\ell_0/a)^2} d \hat{t} \ r_E \ \hat{\rho}_A
  \ \hat{\rho}_B \quad 
\end{equation}
The upper integration boundary is the dimensionless time for
depositing one monolayer.
This approximation can be justified by taking into account that the 
transport of {\em A}-atoms from the $n$th to the $(n+1)$th layer 
mainly takes place at early times, that is, the nucleation regime and 
early stages of 
the intermediate coverage regime, as explained in section \ref{sect:correl}.
The chosen approach describes  these regimes with sufficient accuracy.
  
The solution of the first two equations can be taken directly from 
the literature \cite{Amar94}: 
For early times, $\hat t \ll 1$, $\hat \rho$ is linear in $\hat t$, and 
$\hat I$ increases with $\hat t^3$. At late times, $\hat t \gg 1$, 
one gets $\hat \rho \propto \hat t ^{1/3}$ and
$\hat I \propto \hat t ^{-1/3}$.

With these results, the last two equations can be solved analytically in 
a similar way for the early-time regime, $\hat t \ll 1$.
For equation (\ref{eq:dgl-sys-scale-3}), the second term on the right
hand side can be neglected in this limit. 
If we also neglect the time dependence
of $\hat{\rho}_A(t)  \approx \hat{\rho}_A(0) = \hat c_A(n)$,
we get 
\begin{equation}
  \dot{\hat{\rho}}_B \approx 1 - r_E \ \hat c_A \ \hat{\rho}_B \quad 
  \label{eq:dgl-sys-3-approx}
\end{equation} 
This equation relaxes into a steady state with 
$  \hat \rho_{B,\infty} \propto 1/(r_E \ \hat c_A)$
after a characteristic time 
$\hat t^* \approx 1/(r_E \ \hat c_A(n))$.
For even smaller times, $\hat t \ll \hat t^*$, we can also neglect the 
other right hand side term, and we get $\hat \rho_B \propto \hat t$.

Plugging these results into Eq.(\ref{eq:dgl-sys-scale-4}), and realizing that 
the second term on the right hand side can be neglected
compared to the first one, we get
\begin{equation}
\hat{\rho}_A \propto e^{-\hat t / \hat c_A} \approx ( 1 - \frac{\hat
  t}{\hat c_A}) \qquad \textrm{for}     
  \qquad \hat t^* \ll \hat t \ll 1
\label{eq:rho-a-approx-1}
\end{equation}
and 
\begin{equation}
\hat{\rho}_A \propto e^{- r_E \hat t^2} \approx (1 - r_E \hat t^2)
    \qquad \textrm{for}    \qquad \hat t \ll \hat t^*  \quad  .
\label{eq:rho-a-approx-2}
\end{equation}

To relate these findings to our results 
in the other sections,
we employed an iteration scheme for the rate 
equation system to obtain the surface concentration of $A$'s for every 
deposited  integer monolayer.
Starting from the substrate ($\rho_A(0) = c_A(0) = 1$) we can obtain
$\rho_A(1) = c_A(1)$ from Eq.(\ref{eq:rek-c_A}) by 
solving Eqs.(\ref{eq:dgl-sys-scale-1}) -- (\ref{eq:dgl-sys-scale-4})
numerically.
Plugging $c_A(1)$ back into our rate equations
as the initial surface concentration, 
that is $\rho_A(0) = c_A(1)$, 
we get $\rho_A(1) = c_A(2)$ by using
Eq.(\ref{eq:rek-c_A}) again.
Repeating this iteration scheme leads to $c_A(t)$.

\begin{figure}[ht]
  \begin{center}
    \includegraphics[angle=270, width=85mm]{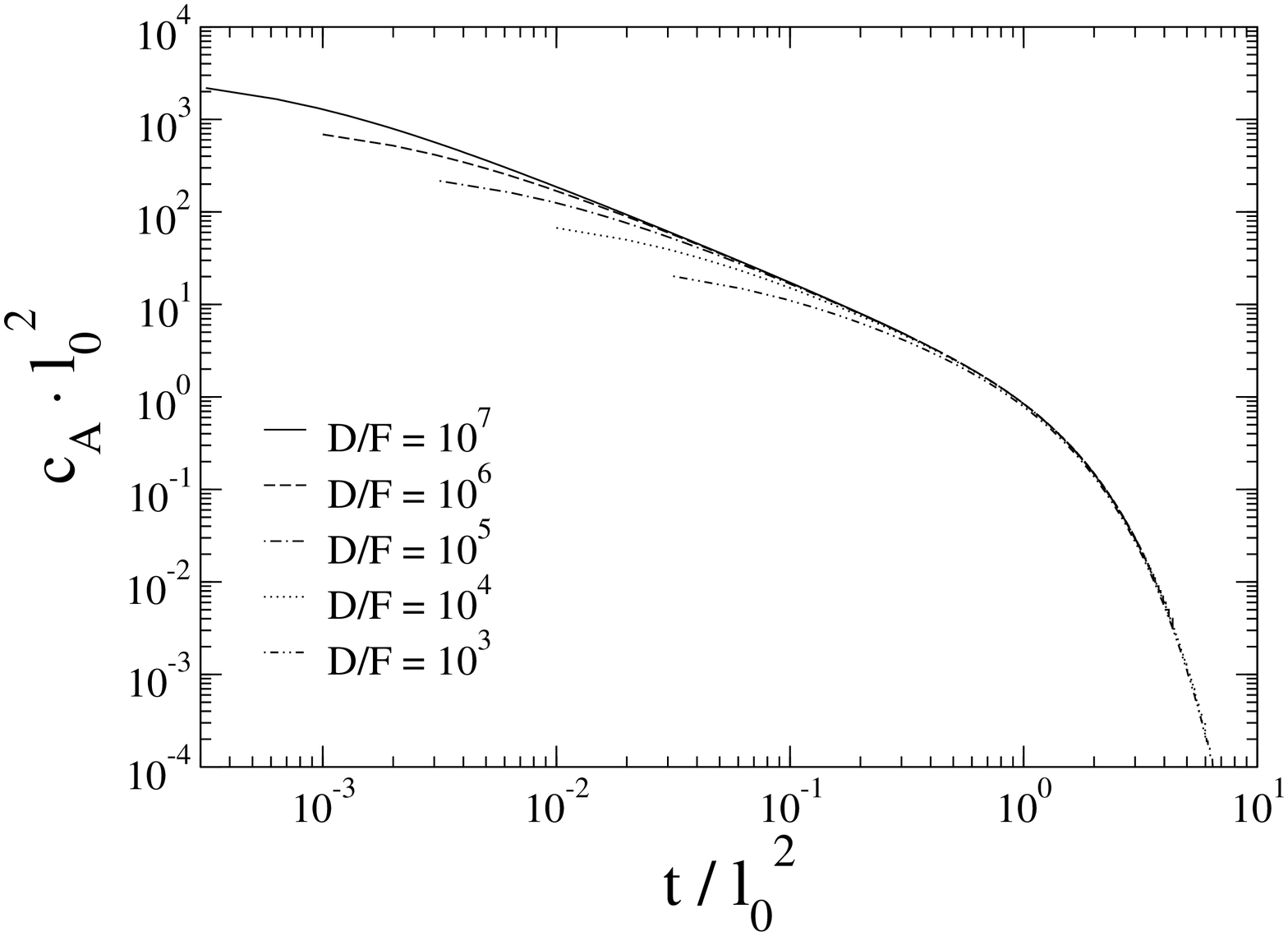}
    \vspace{2mm}
    \smallcaption{
      \label{fig:rate_eq_iter_scale}
      Scaled time dependence of the surface concentration of substrate 
      atoms, obtained from the iteration of the rate equations.      
      The slope of the power law region is $-1.03 \pm 0.05$.
    E/D = 1.}
  \end{center}
\end{figure}

The rescaled results of this approach for different $D/F$ are shown 
in Fig. \ref{fig:rate_eq_iter_scale}.
One can clearly observe power law decay for high $D/F$ values at 
intermediate times,
and a similar scaling behaviour as obtained from the simulations.
The exponent of the power law behaviour is  
approximately $-1$, which is identical to the result from the 
simulations for perfect layer-by-layer growth mode, $D/F \to \infty$.
This fact supports our argumentation concerning the different exponents
for $D/F \to \infty$ and finite $D/F$ in section \ref{finite_dof}:  
Since the rate equations cannot describe the clustering of the
{\em A}-atoms, they also do not reflect the preference of the
nucleation sites to be far from the {\em A}-clusters.
 The observed scaling exponent of $l_0^2 = \left( D/F \right) ^{1/2}$
doesn't match any of the exponents resulting from the simulations.
This is in general agreement with the analytical calculation of the
exponents in section \ref{width}:   
There we derived the scaling exponents from  characteristic properties 
of the {\em A}-clusters, which are totally neglected in 
the presented rate equation approach.

\section{Conclusion}
\label{conclusion}

In the present work we have investigated heteroepitaxial growth of
\bfb-particles on an \bfa-substrate. Introducing an exchange mechanism
for \bfb-adatoms, when they encounter an \bfa-atom in the uppermost
layer, we observed that in the limit of 
layer-by-layer growth the top layer concentration of \bfa-atoms decays
algebraically. Therefore, the resulting interdiffusion zone has
a broad profile with a diverging width. Varying the rate $E$ at which
\bfa-atoms and \bfb-atoms are exchanged did 
not change the exponent of this power law. A different situation has been
observed, as we varied the diffusion constant $D$. For finite values
of $D$ a crossover from power law to exponential decay has been
found. The crossover time $H$ is given by
$(D/F)^{0.3}$ for $d=1$ and by $(D/F)^{0.6}$ for $d=2$. 

It would be very interesting to have some experimental results for
the concentration of substrate atoms in the growing layer in the case
of heteroepitaxy. The results we found suggest that in order to
provide a sharp interface between substrate and deposited material one
should use small $D/F$.

\begin{acknowledgements}
  We are very grateful to H. Hinrichsen, L. Brendel,
  V. Uzdin, U. Koehler, C. Wolf and A. Lorke,
  with whom we had many fruitful discussions. 
  This work was supported by the DFG within 
  Sonderforschungsbereich 491 (Magnetic Heterolayers) and
  GK 277 (Strukture and Dynamics of Heterogeneous Systems), as well as
  by the INTAS-project 2001-0386. Work done by SBL was supported in
  part by Korea Research Foundation Grant (KRF-2001-015-DP0120).
\end{acknowledgements}

\section*{Appendix}
\label{Appendix}  

Here we describe, how we implemented the model introduced in Sec. \ref{sec:model}
for one dimensional surfaces in the limit $D/F \rightarrow \infty$.
We first describe the idea for the scale free limit,
where also $E/D \rightarrow \infty$.

For $D/F \rightarrow \infty$ one has perfect layer-by-layer growth. The 
nucleation of a new layer happens at an arbitrary position. Afterwards 
there is at most one adatom on the surface. The idea is to calculate the 
probabilities exactly, with which the adatom reaches the
nearest sinks to its left and to its right. For an \bfa-adatom these are the 
island edges, while for a 
\bfb-adatom it might also be an \bfa-atom, with which it could exchange.
Let $d_{L}$ ($d_{R}$) denote the distance to the nearest sink to the left 
(right).

As shown in \cite{Gardiner85}, the probability $p_{\rm L}$ 
to reach the left position prior to the right one with unbiased
diffusion is given by
\begin{equation}
  \label{eq:11}
  p_{\rm L}=\frac{d_{\rm R}}{d_{\rm R}+d_{\rm L}}
\end{equation}
Correspondingly, $p_{\rm R}=1-p_{\rm L}$.
Therefore, it is not necessary to simulate the whole random walk of
an adatom, but it suffices to select the final
position according to \eqref{eq:11}.  

Thus in the scale free limit the model (after nucleation of a new
layer) may be simulated as follows:\\ (1) Deposition of a \bfb at a
randomly chosen site $i$.\\ (2) Determination of the distances $d_{\rm
  L}$ and $d_{\rm R}$ followed by a decision for a side according to
the probabilities given in \eqref{eq:11}.\\ (3) If the final position
of the \bfb-adatom is an \bfa-site, the atoms exchange (as $E/D
\rightarrow \infty$). In this case the \bfa-adatom goes to the left or
right island edge according to \eqref{eq:11}. If the final postition
of the \bfb-adatom is an island edge, it is bound there irreversibly
possibly overgrowing an \bfa-atom. Then one returns to step (1) and
deposits the next \bfb-atom.\\ 

This algorithm can be generalized for finite $r_{\rm E}=E/D$: 
Not always, when a 
\bfb-adatom encounters an exchange partner \bfa, they exchange immediately. This
happens only with probability $p_{\rm E}=E/(E+2D)$, where the denominator
is the sum of the rates for the three possible actions of the adatom --
exchange with the A-atom underneath, a hop to the right neighbour and a hop
to the left neighbour. With probability $p_{\rm E}$ the \bfb-adatom is replaced
by an \bfa-adatom, which attaches to the island edges to its left 
with probability Eq.(\ref{eq:11}), and otherwise to the island edge to its right;
$1-p_{\rm E}$ is the probability that the \bfb-adatom
continues to diffuse until it encounters the next \bfa-atom or attaches to the
island edge.

In order to avoid simulating the random walk explicitely, one has to 
calculate the probabilities analytically, that the \bfb-adatom exchanges
with any particular of the \bfa-atoms or attaches to the island edges.
Technically speaking, the \bfb-atom is a random walker on a one-dimensional
lattice with fixed partial absorbers (the \bfa-atoms) and two full absorbers
(the island edges) (Rosenstock trapping model with partial absorbers).
In order to calculate the absorption probabilities at the
different absorbers, which depend on the deposition site, 
we consider an incoming flux (normalized to 1) of independent random
walkers at the deposition site $x_{\rm S}$ (source) and determine the
outgoing fluxes at the 
absorption sites (sinks). The absorption probability is then the
steady state fraction of the incoming flux that leaves the system at the
respective absorption site.  
 
The density of random walkers at a site $x$ evolves according to 
\begin{eqnarray}
\label{diff_react} 
\dot\rho(x,t) &=& D [\rho(x-1,t) - 2\rho(x,t) + \rho(x+1,t)] \\
              & & - E \rho(x,t)\rho_{\rm A}(x)
              + \delta_{x,x_{\rm S}},  \nonumber
\end{eqnarray}
where the density of partial absorbers, $\rho_{\rm A}(x)$, is 1 at
all the sites $x_{\rm A}$, where an \bfa-atom sits, and 0 otherwise:
\begin{equation}   
\label{rho_A}
\rho_{\rm A}(x) = \sum_{\nu=1}^n \delta_{x,x_{{\rm A}\nu}}.
\end{equation}
The terms on the right of Eq.(\ref{diff_react}) which are proportional to $D$ are
the gain and loss terms due to hopping from a neighbor site to $x$,
respectively away from $x$. The term proportional to the exchange rate $E$
describes the loss of walkers at the partial absorption sites.
The last term is the gain term due to the normalized influx of walkers at
site $x_{\rm S}$. The perfect sinks corresponding to the island edges
are represented by the boundary conditions $\rho(1) = \rho(L) = 0$,
where $L$ is the size of the terrace, on which the source is located.

The probability of absorption at site $x_{\rm A}$ is then obtained
from the steady state solution of Eq.(\ref{diff_react}) by
\begin{equation}
\label{px_A}
p(x_{\rm A}) = E \rho(x_{\rm A}),
\end{equation}
and the ones at the island edges by
\begin{equation}
\label{p_edge}
p(1) = D \rho(2), \quad p(L) = D \rho(L-1).
\end{equation}

Introducing the diffusion current between $x$ and $x+1$ (i.e. the current 
to the right of $x$ and to the left of $x+1$),
\begin{equation}
\label{current}
j_{\rm R}(x) = j_{\rm L}(x+1) = - D (\rho(x+1) - \rho(x)),
\end{equation}
equation (\ref{diff_react}) can be rewritten in the steady state as
\begin{equation}
j_{\rm R}(x) - j_{\rm L}(x) = - E \rho(x) \sum_{\nu=1}^n
\delta_{x,x_{{\rm A}\nu}} + \delta_{x,x_{\rm S}}.
\end{equation}
This shows that $\rho(x)$ is a piecewise linear function with 
slope discontinuities at the source and the sinks.
Hence Eq.(\ref{diff_react}) reduces to a set of $2n+2$ coupled linear equations 
for the $2n+2$ unknowns
$j_{\rm R}(x_{\rm A\nu}),   \rho(x_{\rm A\nu})$ and the boundary
values $j_{\rm R}(1)$ and $j_{\rm L}(L)$.

The solution determines the probabilities Eqs.(\ref{px_A}), (\ref{p_edge})
with which a freshly deposited \bfb-atom is exchanged at the different
\bfa-sites or absorbed by the island edges. By choosing a random
number we decide which site to pick.  If it is an island edge, the
\bfb-atom is moved there, and the next \bfb-atom is deposited at a
random position.  Otherwise we move the \bfb-atom to the chosen site,
exchange it with the \bfa-atom there, let another random number
determine, whether to attach the \bfa-atom to the left or right island
boundary, and deposit the next \bfb-atom at a random position.

The complexity of this algorithm is linear in the number of \bfa-atoms left on 
the surface, while a brute force simulation of the diffusion would cost
much more computing time proportional to $L^2$.


\end{document}